\documentclass[a4paper,10pt]{article}
\usepackage{amsmath, amssymb, amscd}
\usepackage[all]{xy}
\begin{document}

\renewcommand{\thefootnote}{\fnsymbol{footnote}}

\begin{titlepage}
\begin{center}
\hfill LTH 877 \\
\vskip 10mm

{\Large
{\bf Non-extremal Black Holes, Harmonic Functions, and 
Attractor Equations}}

\vskip 10mm

\textbf{T. Mohaupt}\footnote{\textsf{Thomas.Mohaupt@liv.ac.uk}}
\textbf{and O. Vaughan}\footnote{\textsf{Owen.Vaughan@liv.ac.uk}}

\vskip 4mm

Theoretical Physics Division\\
Department of Mathematical Sciences\\
University of Liverpool\\
Liverpool L69 7ZL, UK \\

\end{center}

\vskip .2in 

\begin{center} {\bf ABSTRACT} \end{center}
\begin{quotation} \noindent
We present a method which allows to deform extremal
black hole solutions into non-extremal solutions,
for a large class of supersymmetric and non-supersymmetric
Einstein-Vector-Scalar type theories. 
%Using dimensional reduction over time this problem is 
%reformulated in terms of deforming harmonic maps from 
%transverse space into an extended scalar manifold which 
%encodes all relevant degrees of freedom. 
The deformation is shown to be largely 
independent of the details of the matter sector.
While the line element is dressed with an
additional harmonic function, the attractor equations 
for the scalars remain unmodified in suitable coordinates,
and the values of the scalar fields on the outer and inner
horizon are obtained from their fixed point values by
making specific substitutions for the charges.
For a subclass of
models, which includes the five-dimensional 
STU-model, we find explicit
solutions. 
\end{quotation}

\vfill

\end{titlepage}

\eject

\renewcommand{\thefootnote}{\arabic{footnote}}
\setcounter{footnote}{0}

\section{Introduction}

Over the last 15 years there has been tremendous progress 
in understanding the entropy of extremal black holes in 
string theory. While the matching of the microscopic
entropy \cite{Strominger:1996sh} and the macroscopic entropy
\cite{Ferrara:1995ih} of BPS black holes triggered the 
ongoing interest in the subject,  
it has been appreciated more recently that many
features of BPS black holes also apply to non-BPS 
extremal black holes, and, hence do 
not rely critically on 
supersymmetry \cite{Goldstein:2005hq,Tripathy:2005qp}. 
In contrast, progress on non-extremal solutions has
been less impressive. Higher-dimensional non-extremal 
black hole and black brane solution have 
been known for some time, as well as non-extremal solutions of
compactified supergravity theories 
\cite{Tangherlini,MyersPerry,GibbonsMaeda,GarfinkleHorowitzStrominger,HorowitzStrominger,CveticYoum}. 
More recently, it has been observed that various non-extremal solutions
can be obtained by reducing the equations of motion to first
order equations \cite{Lu:2003iv,Miller:2006ay,Garousi:2007zb,Andrianopoli:2007gt,Janssen:2007rc,Cardoso:2008gm,Perz:2008kh}.
Treating near-extremal black holes as composites
of branes and antibranes accounts for the entropy to leading
order, and allows to derive Hawking radiation, including greybody
factors \cite{Callan:1996dv,Horowitz:1996fn,Cvetic:1996kv,Maldacena:1996ix,Behrndt:1997as}.

In this article we develop an approach to 
non-extremal black solutions which keeps the matter sector
as general as possible. Our main focus is to get a systematic
understanding of how extremal solutions can be made non-extremal,
and which features survive this deformation. 
Much of the success
in the study of  extremal black holes is due to the good 
understanding of how they arise as solutions of (super-)gravity in the
presence of a generic matter sector.
Here `generic' means that the matter sector
is as general as allowed by the symmetries underlying 
the action. The attractor mechanism 
\cite{Ferrara:1995ih,Strominger1996,FerraraKallosh} does not only 
guarantee that the near-horizon solution, and, hence, the
entropy is completely determined by the charges,\footnote{Non-BPS attractors
have been studied extensively during the past years, see for example
\cite{Belluci} for a review.}
 but also
allows to find global black hole solutions in terms of harmonic
functions. While solutions cannot always be found in 
completely explicit form, the field equations can be 
reduced to a coupled system of algebraic equations,
sometimes called `generalized stabilization equations',
which express the solution in terms
of harmonic functions \cite{Behrndt:1997ny,LopesCardoso:2000qm}. 
The organization of the solution 
in terms of charges and harmonic functions reflects that from
a higher-dimensional (ten- or eleven-dimensional) point of view,
black holes are composites of branes and other string or M-theory
solitons. This provides the link between black hole thermodynamics
and microscopic properties. 

One well-known feature of black hole and black brane solutions 
in various dimensions is that non-extremal solutions differ from 
extremal ones by the presence of one additional harmonic function,
which parametrizes the deviation from extremality.
We will review this for the five-dimensional
version of the Reissner-Nordstrom solution below. This feature does not
only occur for solutions which carry a single type of charge, 
and thus have a single type of stringy constituent, but also
for more complicated solutions, which are multiply charged and
can be interpreted as composites of various different types of
branes. We interpret this as evidence that the deformation of
extremal into non-extremal solutions is `universal', in the sense
that it is largely blind to details of the matter sector. Establishing
and understanding this in generality is likely to 
enhance our understanding of non-extremal black holes considerably.
In this article we develop an approach 
based on dimensional reduction over time, harmonic maps and generalized
special geometry. Let us explain these key ingredients and compare
them to other approaches taken in the literature. 

Dimensional reduction over time, and, for spherically symmetric
solutions, dimensional reduction to a one-dimensional problem 
involving only the radial variable, is a powerful solution 
generating technique.\footnote{We refer to \cite{Stelle:1998xg} for a review.} 
It has been applied to Kaluza-Klein 
black holes \cite{Breitenlohner:1987dg} and 
brane-type solutions \cite{Clement:1996nh}, while 
in \cite{FKS} dimensional reduction was used 
to obtain the black hole attractor equations 
from the field equations rather than using Killing spinors. 
More recently, this method has been applied more frequently
in the study of extremal non-BPS black holes, and, to some 
extent, non-extremal black holes 
\cite{CerDal,CCDOP,GNPW:05,GaiLiPadi:08,Perz:2008kh,ChiGut,CRTV,ADOT},
and to other brane-type solutions \cite{Bergshoeff:2008be}.  
However, we believe that this method is still under-appreciated, 
and can become even more powerful if the underlying geometry
is fully employed. Dimensional
reduction reduces the field equations to the equations of a 
harmonic map, possibly modified by a potential, 
from the (reduced) space-time into a scalar target
space which encodes all fields contributing to the solution.
For static, spherically symmetric solutions one obtains the
equation for a geodesic curve in the target space, possibly 
modified by a potential. The geometry of the reduced space-time
reflects the ansatz imposed on the unreduced one. In particular,
extremal solutions correspond to flat 
reduced geometries.\footnote{When including Taub-NUT charge, one has to 
consider more general Ricci-flat geometries \cite{Gaiotto:2005gf}.} 
We will see later that in the non-extremal 
case the geometry is the time-reduced version of the simplest
charged non-extremal solution, the Reissner-Nordstrom solution,
independently of the matter content. 
The geometry of the scalar target space encodes the dynamics
of the fields entering into the solution. For supergravity theories
the relevant geometries are symmetric spaces for $N>2$, and
various `special geometries' for $N=2$ supersymmetry. The latter
need not be symmetric or even homogeneous spaces, but are 
characterized by the existence of a potential for the scalar 
metric. As has become clear recently, there is a more general
class of scalar geometries, which might be called `generalized
special geometries', which correspond to non-supersymmetric theories,
and allow the construction of solutions which share the key
features of the solutions of supersymmetric theories \cite{Mohaupt:2009iq}. 
In particular, if one replaces the special real geometry 
of five-dimensional vector multiplets \cite{GST} by the 
`generalized special real geometry' introduced in \cite{Mohaupt:2009iq},
then the attractor equations still have the same form 
discovered in \cite{Sabra5d,ChaSab} for five-dimensional supergravity, 
and extremal multi-centered 
solutions can be obtained in terms of harmonic functions.

In this article we apply this type of approach to the construction 
of non-extremal solutions. We restrict ourselves to static, spherically
symmetric solutions for simplicity. As in \cite{Mohaupt:2009iq} 
we impose that the
scalar geometry of the underlying theory, before dimensional reduction,
is `generalized special real', and for concreteness we start
from five dimensions. This is natural, because generalized special
real geometry is a generalization of the special real geometry
of five-dimensional vector multiplets. As supersymmetry
does not play a role, our results could easily be adapted to any
dimension $d\geq 4$ by adjusting numerical parameters.\footnote{The formulae 
we use for dimensional reduction contain parameters whose values depend on 
the number of space-time dimensions. We felt that it was too cumbersome to
include this depdendence throughout.} One limitation which we need 
to mention is that we only obtain black hole solutions with electric 
charges. While this is no restriction in $d>4$, in $d=4$
charged black holes can carry both electric and magnetic charge. 
There is no probem in principle with applying temporal reduction 
to a four-dimensional theory, but, as is well known from 
the $c$-map \cite{FerSab},
the isometry group of the resulting scalar manifold is more
complicated. Instead of the abelian groups occurring in this paper
one obtains solvable Lie groups (of Heisenberg group type). 
This appears to be a technical rather than conceptual complication, 
and we have  decided to consider the simpler case of abelian isometry groups
in this paper, while dyonic solutions are left to future work.

As in \cite{Mohaupt:2009iq} our strategy is to simplify the equations of motion
until the solution can be expressed in terms of harmonic functions.
This is similar in spirit to the way the `generalized stabilization 
equations' are derived in the framework of the superconformal 
calculus \cite{LopesCardoso:2000qm}. 
An alternative approach is to reduce the equations of 
motion to first order form, leading to gradient flow equations
\cite{Miller:2006ay, CerDal,CCDOP,Janssen:2007rc,Cardoso:2008gm,Perz:2008kh}.
This approach mimics the Killing spinor equations of BPS solutions,
with the central charge being replaced by a `fake superpotential'
which drives the flow. In our approach the re-writing of the 
field equations in first order form is sidestepped, so that we
obtain the solution directly. For the extremal case it was 
explained in \cite{Mohaupt:2009iq} how to obtain the flow equations starting
from the harmonic map equation. We expect that this relation can
be generalied to cover the results obtained for non-extremal 
solutions in this paper, but leave a detailed investigation to
future work.

Many of the results obtained in the literature are based on the
assumption that the scalar target is a symmetric space and exploit
the relation to integrability and the Hamilton-Jacobi formalism
\cite{GNPW:05,GaiLiPadi:08,Perz:2008kh,ChiGut,CRTV,ADOT}. Our 
approach attempts to be less restrictive and only requires the 
scalar metric to have a potential. Thus roughly speaking we try
to work in the analogue of an `$N=2$ framework' (special geometry, 
prepotentials)
rather than an `$N>2$ framework' (symmetric spaces, integrability). 
While the explicit non-extremal solutions
obtained in this paper happen to correspond to symmetric targets, we argue
that the structures which we discover hold more generally, and that
the method we are developing is general and flexible enough to 
deal with target manifolds which are not symmeric spaces.
This is supported by the previous observation that  
extremal multi-centered solutions can be  
constructed easily for the whole class of models based on generalized 
special real geometry \cite{Mohaupt:2009iq}. 
Of course, symmetric spaces provide an important and interesting 
class, and the relation between our approach and the one based on 
integrability should be clarified in the future.

This paper is organised as follows. In Section 2 we first
review the five-dimensional version of the Reissner-Nordstrom 
solution. Then we perform the reduction of a five-dimensional
action based on generalized special real geometry, first
with respect to time, then, assuming spherical symmetry, to 
a one-dimensional effective theory of the radial degrees of freedom.
We make some observations which are very helpful in the following:
the geometry obtained after reduction over time is, when assuming 
spherical symmetry, the time-reduced five-dimensional Reissner-Nordstrom
metric, irrespective of the matter content. We also identify two
useful radial coordinates: the affine curve parameter 
$\tau$, which is only defined outside the 
outer horizon, and the isotropic radial coordinate 
$\rho$, which allows us to extend solutions 
up to the inner horizon. After reviewing the relevant background 
material about generalized special real geometry, we analyze and
simplify the remaining equations of motion. We identify a subclass
of models, dubbed `diagonal', where solutions can be obtained in 
closed form. Finding explicit solutions for more general models is
left to future work. In Section 3 we lift our solutions to five
dimensions and investigate their properties. For diagonal models
we obtain non-extremal solutions, valid up to the inner horizon,
where all scalar fields are non-constant. The solutions are given 
in terms of harmonic functions, with one particular function 
encoding the non-extremality. Extremal solutions are related
to non-extremal solutions with the same charges by 
dressing them in a specific way with the additional harmonic
function. In a particular parametrization the expressions for 
the five-dimensional scalars are identical to the extremal 
case and solve the same generalized stabilization equations.
While there is no attractor or fixed point behaviour in the
proper sense, the values of the scalars on the outer and
inner horizon are obtained from the fixed point values by
specific substitutions, which replace charges by `dressed' 
charges. Then we turn to a particular diagonal model, the
five-dimensional STU-model, which can be obtained (as a subsector)
by compactification of type-IIB string theory on $T^4 \times S^1$.
We show how our solution is related to the D5--D1 system, and
thus establish the relation between our charge parameters and 
the microscopic charges corresponding to D-branes. Then we
turn to the universal solution, which exists in all our models,
and show that all five-dimensional scalars are constant, while
the metric is the five-dimensional Reissner-Nordstrom metric. 
Following this we briefly comment on `block-diagonal' models,
where the scalar manifold is a product. In this case we 
obtain solutions where some, but not all scalars can be
non-constant. In Section 4 we discuss our results and give 
an outlook on future research.

\section{Dimensional reduction and instanton solutions}

\subsection{Review of the five-dimensional Reissner-Nordstrom 
black hole}

Some clues how non-extremal, static, spherically symmetric 
solutions 
should be approached within the setting of
dimensional reduction, harmonic maps, and generalized special geometry 
can be taken
from the five-dimensional version of the 
Reissner-Nordstrom solution. One standard form
of the line element is \cite{Tangherlini,MyersPerry}
\[
ds_{(5)}^2 = - \frac{(r^2 - r_+^2)(r^2 - r_-^2)}{r^4} dt^2 
+ \left[ \frac{(r^2 - r_+^2)(r^2 - r_-^2)}{r^4} \right]^{-1} dr^2 
+ r^2 d \Omega^2_{(3)}  \;.
\]
In this coordinate system the singularity is located at the origin,
$r=0$, whereas $r_- >0$ is the inner horizon (Cauchy horizon) and 
$r_+>r_-$ is the outer horizon (event horizon). In the extremal limit
both horizons coincide, $r_+=r_-$. Deviations from extremality can 
be parametrized by the non-extremality parameter $c= \frac{1}{2} 
(r_+^2 - r_-^2) \geq 0$. For the construction of black hole and black brane
solutions one often prefers isotropic coordinates, in which the
spatial part of the metric is conformally flat. For the five-dimensional
Reissner-Nordstrom solution this is achieved by introducing the new
radial coordinate $\rho$, where
\[
\rho^2 = r^2 - r_-^2 \;.
\]
This coordinate system is centered at the inner horizon, which is 
at $\rho=0$, while the outer horizon is at $\rho^2 = 2c$. In isotropic
coordinates the line element takes the form
\begin{equation}
\label{5dRN}
ds_{(5)}^2= -\frac{W}{ {\cal H}^2 } dt^2 + {\cal H} \left[ W^{-1} d\rho^2 
+ \rho^2 d\Omega^2_{(3)} \right] \;.
\end{equation}
which is parametrized in terms of
two harmonic functions\footnote{Here and in the following `harmonic
function' refers to a function which is harmonic in the 
coordinates transverse
to the worldline of the black holes (i.e., the four spatial coordinates), 
with respect to the standard, `flat' Laplacian.}
\[
{\cal H} = 1 + \frac{q}{\rho^2} \;,\;\;\;
W = 1 - \frac{2c}{\rho^2} \;.
\]
The parameter $q$, which is the electric charge carried by the 
black hole\footnote{Acutally, $q$ is the modulus of the electric charge.
Observe that $q$ cannot be negative, as this would introduce additional
singularities in the line element. Note that since the energy momentum tensor is
quadratic in the Maxwell field strength, the Einstein equations do not `see'
the sign of the charge. For convenience, we will refer to $q$ as the
electric charge.}
is related to $r_-$ by 
$q:=r_-^2$. We prefer to parametrize black holes solutions by the
electric charge $q$ and the non-extremality parameter $c$ instead
of the positions $r_\pm$ of the horizons.

Two interesting limits can be obtained by 
switching off either of these `charges'. Setting $q=0$ we obtain 
a five-dimensional version of the Schwarzschild solution, while
setting $c=0$ makes the solution extremal. Thus deforming 
the solution away from extremality amounts to `switching on' 
an additional harmonic function in the line element.
Experience with supersymmetric solitons in various dimensions
suggests that this is a generic feature. 

If we perform a dimensional reduction with respect to time,
then the four-dimensional (Einstein frame) metric
$ds^2_{(4)}$ is related to the five-dimensional 
(Einstein frame) metric by
\begin{equation}
\label{5dKK4d}
ds_{(5)}^2 = -e^{2\tilde{\sigma}}dt^2 + e^{-\tilde{\sigma}} ds^2_{(4)}\;.	 
\end{equation}
For the five-dimensional Reissner-Nordstrom solution 
the Kaluza-Klein scalar $\tilde{\sigma}$ is given by
\[
e^{2 \tilde{\sigma}} = \frac{W}{{\cal H}^2}\;.
\]
The extremal limit ($W=1$) has the particular feature
that the reduced line element $ds^2_{(4)}$ is flat. As we will
see in more detail below,  
constructing extremal black hole solutions therefore amounts to constructing
a harmonic map from a flat 
manifold (reduced space-time) into a scalar target space, which 
in Einstein-Maxwell theory accomodates the Kaluza-Klein scalar and
the electro-static potential. The solution corresponds to a null
geodesic curve in the scalar target space. 
Once we consider non-extremal solutions, where  $W\not=1$, the 
reduced space-time metric $ds^2_{(4)}$ is no longer flat, and the
geodesic curve in the scalar target space is no longer null.
Our main strategy is to disentangle the non-extremal deformation,
which is encoded in the additional harmonic function $W$, from 
the degrees of freedom already present in the extremal case.

\subsection{Dimensional reduction}

We begin by considering a five-dimensional 
action of scalars and abelian gauge fields 
coupled to gravity. 
\begin{equation}
\hat{S} = \frac{1}{8 \pi G_N^{(5)}} \int d^5 \hat{x} \sqrt{| \hat{g} |} 
\left[ \frac{\hat{R}}{2} - \frac{3}{4} a_{IJ}(h)	\partial_{\hat{\mu}} h^I \partial^{\hat{\mu}} h^J 
- \frac{1}{4} a_{IJ}(h) \hat{F}^I_{\hat{\mu} \hat{\nu}} \hat{F}^{J \hat{\mu} \hat{\nu}} + \ldots \right] \;,
\label{5d_action}
\end{equation}
where $I = 1, \ldots, n$ and $\hat{F}^I_{\hat{\mu} \hat{\nu}} = \partial_{\hat{\mu}} \hat{\mathcal{A}}^I_{\hat{\nu}} - \partial_{\hat{\nu}} \hat{\mathcal{A}}^I_{\hat{\mu}}$.

The dots represent further terms like Chern-Simons and fermionic terms,
which could be present, but do not contribute to backgrounds which are
static and purely electric. The truncation of five-dimensional supergravity
coupled to $n-1$ vector multiplets to such a background
has the above form, with a `special real' scalar metric $a_{IJ}$. 
This means that the metric has a Hesse potential ${\cal V}(h)$,
\[
a_{IJ}(h) = \partial_I \partial_J {\cal V}(h) \;,
\]
and where the Hesse potential takes the special form 
${\cal V}(h) = -\log{\cal \hat{V}}(h)$, with a  `prepotenial'
${\cal \hat{V}}(h)$ which is a homogeneous cubic polynomial. In addition, 
the scalars must satisfy the hypersurface constraint
\begin{equation}
{\cal \hat{V}}(h) = 1 \;.
\label{hypersurface_constraint}
\end{equation} 
This means that the manifold parametrized by the physical scalar fields
is a hypersurface 
$\hat{M} = \{ \hat{\cal V}(h) = 1 \}$ 
in a Hessian manifold $M$ with metric
$a_{IJ}$. The metric on the hypersurface $\hat{M}$ is the pull-back of 
$a_{IJ}$. We will not limit ourselves to supersymmetric theories and 
allow a larger class of scalar metrics, where the prepotential 
$\hat{\cal V}(h)$ is a homogeneous function of arbitrary degree $p$.
Such manifolds might be called `generalized special real manifolds',
as they are  natural generalizations of the scalar manifolds occuring 
in supersymmetric theories. The relevant properties of Hessian 
and (generalized) special real manifolds will be presented in the
next section. 

We are only interested in five-dimensional 
solutions which are static and purely electric. 
In order to construct these solutions we perform a time-like dimensional 
reduction where we decompose the metric and gauge vectors as 
follows:\footnote{More details can be found in \cite{EucIII,Mohaupt:2009iq}.}
\[
\hat{g} = \left( \begin{array}{c|c} -e^{2 \tilde{\sigma}} &	-e^{2 \tilde{\sigma}} \mathcal{A}_\nu	\\ \hline & \\-e^{2 \tilde{\sigma}} \mathcal{A}_\mu	& e^{-\tilde{\sigma}} \left( g_{\mu \nu} -e^{2 \tilde{\sigma}} \mathcal{A}_\mu \mathcal{A}_\nu \right) \\
& \end{array} \right) \;,\;\;\;\;\;\;
\hat{\mathcal{A}^I} = \left( \begin{array}{c}	\mathcal{A}^I_0 \\ \hline \\
\mathcal{A}^I_\mu + \mathcal{A}^I_0 \mathcal{A}_\mu \\ \\ \end{array}	\right) \;.
\]
For our class of solutions the Kaluza Klein-vector 
$\mathcal{A}_\mu$ vanishes and the 
last term in the 
Lagrangian becomes
\[
\hat{F}^I_{\hat{a} \hat{b}} \hat{F}^{J \hat{a} \hat{b}} = - 2e^{-2 \tilde{\sigma}} \partial_\mu m^I \partial^\mu m^J \;,
\]
where we have made the identification $m^I = \mathcal{A}^I_0$. 
The resulting four dimensional Euclidean action is
\begin{eqnarray}
S &=& \frac{1}{8 \pi G_N^{(4)}} \int d^4 x \sqrt{|g|} \left[ \frac{R}{2} - \frac{3}{4} \partial_\mu \tilde{\sigma} \partial^\mu \tilde{\sigma}
- \frac{3}{4}a_{IJ}(h) \partial_\mu h^I \partial^\nu h^J \right. \nonumber \\
&&
+ \left.\frac{1}{2} e^{-2 \tilde{\sigma}} a_{IJ}(h) \partial_\mu m^I \partial^\mu m^J  
+ \cdots \right] \;. 
\end{eqnarray}
As indicated we neglect terms that will not contribute to the type 
of solution we are interested in. In particular, we neglect four-dimensional
gauge fields, because they descend from the magnetic components of the
five-dimensional gauge fields.	
Following the procedure in \cite{Mohaupt:2009iq} we make the rescalings 
\begin{equation}
h^I = e^{- \tilde{\sigma}} \sigma^I \;,\;\;\; 
m^I = \pm \sqrt{ \frac{3}{2} }  b^I \;,\label{5dScalars4d}
\end{equation}
in order to write the action in the convenient form
\begin{equation}
\label{RescaledAction}
S = \frac{1}{8 \pi G_N^{(4)}} \int d^4 x \sqrt{|g|} \left[ \frac{R}{2} - \frac{3}{4}a_{IJ} (\sigma) \left( \partial_\mu \sigma^I \partial^\mu \sigma^J
- \partial_\mu b^I \partial^\mu b^J \right) \right] \;,
\end{equation}
where we have set $a_{IJ}(\sigma) = e^{-2\tilde{\sigma}}a_{IJ}(h)$ using that  
$a_{IJ}$ is homogeneous of degree $-2$. Similarly, we have
\begin{equation}
{\cal \hat{V}}(\sigma) = e^{p \tilde{\sigma}} {\cal \hat{V}}(h) = e^{p \tilde{\sigma}} \;,
\label{KK_prepotential_relation}
\end{equation}
since the prepotential is homogeneous of degree $p$.

Note that while the scalars $h^I$ are subject to the constraint
(\ref{hypersurface_constraint}), the scalars $\sigma^I$ are 
unconstrained and combine the $(n-1)$ five-dimensional scalars with 
the Kaluza-Klein scalar $\tilde{\sigma}$. 
The scalars $\sigma^I$ can be interpreted as
affine coordinates on an $n$-dimensional manifold $M$ with 
Hessian metric $a_{IJ}(\sigma)$. The scalar manifold of the 
five-dimensional theory is embedded into $M$ as a homogeneous
hypersurface $\hat{M}$. In addition to the $\sigma^I$, the four-dimensional
theory has $n$ further scalar fields $b^I$, which descend from the
five-dimensional gauge fields. The gauge symmetries of the five-dimensional
theory induce $n$-commuting isometries $b^I \to b^I + C^I$.
The resulting $2n$ scalar manifold $N$ of the four-dimensional theory 
can therefore be interpreted as the tangent bundle $N=TM$ of $M$. 
The Hessian metric of $M$ extends to a split-signature Riemannian
metric $a_{IJ}(\sigma) \oplus (-1) a_{IJ}(\sigma)$ on $N$. It is easy
to see that this is a para-K\"ahler metric\footnote{We refer to \cite{EucI,EucIII} for a detailed account of para-K\"ahler geometry.} 
and that the Hesse potential
of $M$ is a para-K\"ahler potential for $N$ \cite{Mohaupt:2009iq}.

The four-dimensional equations of motion are
\begin{equation}
\frac{1}{\sqrt{|g|}} \partial^\mu \left( \sqrt{|g|} a_{IJ}(\sigma) \partial_\mu \sigma^J \right)
-	\frac{1}{2} \partial_I a_{JK} \left( \partial_\mu \sigma^J \partial^\mu \sigma^K - \partial_\mu b^J \partial^\mu b^K \right) = 0 \;,
\label{eom1}
\end{equation}
\begin{equation}
\partial^\mu \left( \sqrt{|g|} a_{IJ}(\sigma) \partial_\mu b^J \right) = 0 \;,
\label{eom2}
\end{equation}
\begin{eqnarray}
\frac{1}{4} a_{IJ}(\sigma) \left(\partial_\mu \sigma^I \partial_\nu \sigma^J - \partial_\mu b^I \partial_\nu b^J \right) 
-\frac{1}{8} a_{IJ}(\sigma) g_{\mu \nu} \left(\partial_\gamma \sigma^I \partial^\gamma \sigma^J - \partial_\gamma b^I\partial^\gamma b^J \right)  & & \nonumber \\
= \frac{1}{6} R_{\mu \nu} - \frac{1}{12} R g_{\mu \nu} \;. & & 
\label{eom3_full}
\end{eqnarray}
The first two equations are the scalar equations of motion. They
are equivalent to the geometrical 
statement that critical points of the action with respect to variation of 
$(\sigma^I, b^I)$ define a harmonic map from four-dimensional space-`time'
(with positive definite metric $g_{\mu \nu}$) into the scalar 
target manifold $N$ with metric $a_{IJ} \oplus (-1) a_{IJ}$. The third
set of equations are Einstein's equations. They can be simplified
by taking the trace of (\ref{eom3_full}) and re-substituting the result back: 
\begin{equation}
\frac{1}{4} a_{IJ}(\sigma) \left(\partial_\mu \sigma^I \partial_\nu \sigma^J - \partial_\mu b^I \partial_\nu 	
b^J \right) = \frac{1}{6} R_{\mu \nu} \;.
\label{eom3}
\end{equation}

%%%%%%%%%%%%%%%%%%%%%%%%%%%%%%%%%%%%%%%%%%%%%%%%%%%%%%%%%%%%%%%%%%%%%%%%%%%%
%%%No further discussion of non-spherical solutions in short paper.
%%%%%%%%%%%%%%%%%%%%%%%%%%%%%%%%%%%%%%%%%%%%%%%%%%%%%%%%%%%%%%%%%%%%%%%%%%%%

We now impose that the solution is spherically symmetric.\footnote{This type
of  reduction is frequently used in the literature, see in particular 
\cite{FKS,Perz:2008kh}.} 
A general spherically symmetric line element can be written in the form \cite{Perz:2008kh}
\begin{equation}
ds^2_{(4)} = e^{6A(\tau)} d\tau^2 + e^{2A(\tau)} d\Omega^2_{(3)} \;,
\label{eq:general_spherical_metric}
\end{equation}
where $\tau$ is a radial coordinate. 
The advantage of this parametrization becomes apparent once we look at
the reduced equations of motions for the scalar fields:
\begin{equation}
\frac{d}{d\tau} \left( a_{IJ}(\sigma) \dot{\sigma}^J \right) - \frac{1}{2} \partial_I a_{JK}(\sigma) \left( \dot{\sigma}^J \dot{\sigma}^K - \dot{b}^J \dot{b}^K \right) = 0 \;,\label{sig_model_eom1}
\end{equation}
\begin{equation}
\frac{d}{d\tau} \left( a_{IJ}(\sigma) \dot{b}^J \right) = 0 \;.
\label{sig_model_eom2}
\end{equation}
%\begin{equation}
%\frac{1}{4}a_{IJ}(\sigma) \left( \dot{\sigma}^I \dot{\sigma}^J - \dot{b}^I \dot{b}^J \right) = c^2 \;. \label{sig_model_eom3}
%\end{equation}
These are the equations for a geodesic curve on $N$, written in 
terms of the coordinates $(\sigma^I,b^I)$. For a harmonic map
defined on a one-dimensional domain 
the harmonic equation and the geodesic equation 
coincide.\footnote{In general, the harmonic  equation is the trace of
the geodesic equation and therefore a weaker condition.} We observe
that the geodesic equation is in {\em affine form}, which shows
that the radial coordinate $\tau$ is an affine curve parameter. 
Other parametrizations of the four-dimensional line element use
radial coordinates which are non-affine curve parameters.
The reason for $\tau$ being an affine parameter is that
the Laplace operator for a line 
element of the form (\ref{eq:general_spherical_metric})
takes the form $\Delta = \frac{\partial^2}{\partial \tau^2} + $
terms independent of $\tau$. 

The equations (\ref{sig_model_eom1}) and
(\ref{sig_model_eom2}) follow from the variation of the effective action 
\begin{equation}
S_{eff} = \int d\tau \frac{1}{4}a_{IJ}(\sigma) \left( \dot{\sigma}^I \dot{\sigma}^J - \dot{b}^I \dot{b}^J \right) \;,
\label{1Daction}
\end{equation}
which is the reduction of (\ref{RescaledAction}) in the spherically
symmetric background (\ref{eq:general_spherical_metric}).

We still have to reduce the Einstein equations (\ref{eom3}).
Since we impose spherical symmetry on the scalar fields, 
the LHS of (\ref{eom3}), which is essentially energy momentum 
tensor, vanishes for all components with $\mu,\nu\not=\tau$. 
The corresponding components of the Ricci tensor on the RHS
of (\ref{eom3}) are proportional to 
$\ddot{A} - 2e^{4A}$, and therefore the Einstein equations imply 
\begin{equation}
\ddot{A} - 2e^{4A} = 0. \label{on_shell_A1}
\end{equation}

We now consider (\ref{eom3}) when $\mu = \nu = \tau$. In this case
\begin{equation}
\frac{1}{4} a_{IJ}(\sigma) \left(\dot{\sigma}^I \dot{\sigma}^J - \dot{b}^I \dot{b}^J \right)
=\dot{A}^2 - \frac{1}{2}\ddot{A} = c^2 \;, \label{on_shell_A2}
\end{equation}
where $c^2$ is a constant, which we will choose positive below. 
The fact that  $\dot{A}^2 - \frac{1}{2} \ddot{A}$ must be constant
follows from  (\ref{on_shell_A1}). 
We can combine (\ref{on_shell_A1}) and (\ref{on_shell_A2}) to get
\begin{equation}
\dot{A}^2 = c^2 + e^{4A} \;.
\end{equation} 
This first order equation can be solved as follows, for positive $c^2$:
Taking the square root and multiplying by $-2e^{-2A}$ we find
\[ -2\dot{A}e^{-2A} = \pm2\sqrt{c^2e^{-4A} + 1} \;. \]
We can then relabel $y(\tau) = e^{-2A(\tau)}$ and hence the equation becomes $\dot{y} = \pm2\sqrt{c^2y^2 + 1}$. Solving this we find 
\[ y(\tau) = \frac{\sinh(\pm2c\tau + D)}{c} \;. \]
To ensure $y(\tau)$ is positive we choose the positive sign and $D = 0$. 
We also observe that a negative $c^2$ would lead to 
an equation which is solved by trigonometric rather than hyperbolic
functions. The resulting solutions are periodic in the radial coordinate
and therefore not asymptotically flat. We discard them because we 
want to construct five-dimensional black holes solutions.\footnote{If the 
radial coordinate
is analytically continued and becomes timelike, such solutions might 
correspond to 
cyclic cosmological solutions.}  

Thus we find $e^{-2A} = \frac{1}{c}\sinh(2c\tau)$ and our line element 
is
\begin{equation}
ds_{(4)}^2 = \frac{ c^3 }{ \sinh^3(2c \tau) } d \tau^2  + \frac{c }{ \sinh(2c \tau) } d \Omega^2_{(3)} \;. \label{4D_RN_metric}
\end{equation}
To see that this is in fact the time reduced Reissner-Nordstrom
metric, we replace $\tau$ by a new radial coordinate $\rho$,
which is defined by 
\begin{equation}
%r^2 - r_-^2 = 
\rho^2 = \frac{c e^{2c\tau}}{\sinh(2c\tau)} \;. \label{eq:tau_r_relation}
\end{equation}
Using this new coordinate, the line element takes the form 
\begin{equation}
\label{RN4d}
ds^2_{(4)} = W^{-\frac{1}{2}} d\rho^2 + W^\frac{1}{2} \rho^2 d\Omega^2_{(3)} 
\;, 
\end{equation}
where 
\begin{equation}
\label{Wrhotau}
W = 1 - \frac{2c}{\rho^2}  = e^{-4c\tau}\;.
\end{equation}
To see that this is the time reduced Reissner-Nordstrom metric, we
compare the five-dimensional Reissner-Nordstrom metric (\ref{5dRN}) 
to the Kaluza-Klein ansatz (\ref{5dKK4d}) which relates the five-dimensional 
to the four-dimensional Einstein frame, and observe that the 
resulting Euclidean four-dimensional line element is (\ref{RN4d}). 
We note that the four-dimensional metric takes this form irrespective
of the scalar sector. 

From (\ref{Wrhotau}) it is manifest that the coordinate 
$\tau$ with range $0 < \tau < \infty$ only covers the
range of $\rho$ where $\rho^2>2c$. For 
$0 < \rho^2 < 2c$ the line element (\ref{RN4d}) becomes 
imaginary, but looking back at (\ref{5dRN}) we see that
the five-dimensional line element obtained by lifting is real,
and that $0<\rho^2<2c$ corresponds to the region between the 
outer (event) and the inner (Cauchy) horizon. In this region
the coordinate $t$ becomes space-like while $\rho$ becomes
space-like.\footnote{To be precise, $\rho$ can be continued analytically
beyond the event horizon, while $t$ cannot. However, one can introduce a 
space-like coordinate (which is not the analytical continuation of the 
coordinate $t$ 
used outside the horizon), such that the line element takes the 
form (\ref{5dRN}) between the outer and the inner horizon \cite{HawEll}.}
It is not surprising that our method,
which is based on dimensional reduction over time, does a priori
only give us a solution valid outside the event horizon. 
However, after replacing $\tau$ by $\rho$ the 
analytical continuation to $0<\rho^2<2c$ gives
the Reissner-Nordstrom solution up to the inner horizon. Since
we have seen that (\ref{RN4d}) remains unchanged when
admitting a more complicated matter sector,
we should expect that a similar extension is possible in the
presence of non-constant scalar fields. We will come back to this
later.

The four-dimensional
Einstein equations require that the scalar fields satisfy 
\begin{equation}
\frac{1}{4}a_{IJ}(\sigma) \left( \dot{\sigma}^I \dot{\sigma}^J - 
\dot{b}^I \dot{b}^J \right) = c^2 \;. \label{sig_model_eom3}
\end{equation}
This equation does not follow from the reduced action (\ref{1Daction}),
and must be imposed as a constraint. (It is often called the Hamiltonian
constraint, because it descends from the Einstein equations, which 
are constraints in the Hamiltonian formalism.) Geometrically 
(\ref{sig_model_eom3}) imposes that the norm of the geodesic 
vector field $(\sigma^I, b^I)$ is constant, 
and is given by the parameter $c$ which appears
in the space-time metric. This equation is consistent  
with (\ref{sig_model_eom1}) and (\ref{sig_model_eom2}),
because $\tau$ is an affine curve parameter.\footnote{Affine curve parameters
are singled out by imposing that the norm of the tangent vector is constant 
along the curve. This is necessary and sufficient for the geodesic 
equation to take affine form.}

While the four-dimensional line element is universal, in the
sense that it is independent of the scalar sector, 
the five-dimensional line element depends on the solution
of the scalar field equations through the Kaluza-Klein 
scalar $\tilde{\sigma}$, which is determined by the 
four-dimensional scalars through (\ref{KK_prepotential_relation}).
In particular, if the resulting five-dimensional scalars
are not constant, then the five-dimensional line element 
will be different from the five-dimensional Reissner-Nordstrom
metric.

We remark that it is very encouraging that the four-dimensional metric is
completely determined, and equal to the time-reduced Reissner
Nordstrom metric, irrespective of the matter content of the
theory. This supports the idea that the deformation of 
extremal into non-extremal solutions has universal features and
can be understood in generality, for `arbitrary' matter content.
All features of the solution which depend on the matter sector
are encoded in the Kaluza-Klein scalar which is determined by
the four-dimensional scalar field equations.
Non-extremal solutions
differ from extremal solutions through the replacement of the
four-dimensional flat metric by the time-reduced Reissner-Nordstrom metric,
which is parametrized by a single additional parameter $c$.
Therefore it is reasonable to expect that there is a canonical
one-parameter deformation of the harmonic map corresponding to
an extremal solution, which deforms a null geodesic in $N$ into
a space-like geodesic. This deformation is induced by 
the deformation of the metric on the domain of the 
harmonic map from a flat metric to the time-reduced Reissner-Nordstrom
metric.

%%%%%%%%%%%%%%%%%%%%%%%%%%%%%%%%%%%%%%%%%%%%%%%%%%%%%%%%%

\subsection{Hessian manifolds and dual coordinates}

In order to solve the remaining equations, 
we will use the special geometric properties of the target manifold
$N=TM$. Since $N$ is completely determined by $M$, the essential
properties are those of the Hessian metric $a_{IJ}(\sigma)$ of $M$.
We now collect the relevant properties of Hessian and
(generalized) special real metrics \cite{EucIII,Mohaupt:2009iq}.

A Hessian manifold $(M,a,\nabla)$ is a manifold $M$ equipped
with a pseudo-Riemannian metric $a$ and a flat, torsion-free 
connection $\nabla$, such that the third rank tensor $\nabla a$ is 
completely symmetric.\footnote{The connection $\nabla$ is in general different
from the Levi-Civita connection.} In affine coordinates $\sigma^I$, 
where $\nabla_I = \partial_I$, this
is equivalent to the statement that $\partial_I a_{JK}$ is
completely symmetric. This is the integrability condition
for the existence of a Hesse potential for the metric. Thus
an equivalent local definition in terms of affine coordinates is
that the metric can be written in the form
\begin{equation}
a_{IJ}(\sigma) = \partial_I \partial_J {\cal V} = {\cal V}_{IJ} \;,
\label{hessian_metric}
\end{equation}
where we have introduced the notation $\partial_I {\cal V} = {\cal V}_I, 
\ldots$. 
In affine coordinates, the Christoffel symbols of the first kind 
are completely symmetric and proportional to the third derivatives
of the Hesse potential.

For a (generalized) special real metric we impose in addition 
that the Hesse potential ${\cal V}$ has the form 
\begin{equation}
{\cal V} = - \frac{1}{p} \log {\cal \hat{V}(\sigma)} \;,
\label{hesse_prepotential}
\end{equation}
where the `prepotential' ${\cal \hat{V}}$ is a homogeneous function 
of degree $p$:\footnote{For the special real metrics of five-dimensional
supersymmetric theories, $p=3$, and $\hat{\cal V}$ must be a polynomial.}
\begin{equation}
{\cal \hat{V}}(\lambda \sigma^1, \ldots, \lambda \sigma^n) = \lambda^p {\cal \hat{V}}(\sigma^1, \ldots, \sigma^n) \;.
\label{homogeneous_prepotential}
\end{equation}
It was shown in \cite{Mohaupt:2009iq} that 
Hesse potentials of this form define four-dimensional 
models which can be lifted consistently 
to five-dimensional Einstein-Maxwell-Scalar type theories such 
as (\ref{5d_action}).

Using the homogeneity of the prepotential we deduce that
\begin{equation}
{\cal \hat{V}}_I (\sigma) \sigma^I = p {\cal \hat{V}} (\sigma) \;, \label{hesse_identity1}
\end{equation}
and differentiation implies
\begin{equation}
{\cal \hat{V}}_{IJ} \sigma^I = (p - 1) {\cal \hat{V}}_J  \;. \label{hesse_identity2}
\end{equation}
If we write the metric in terms of the prepotential 
\begin{equation}
a_{IJ}(\sigma) = {\cal V}_{IJ} = - \frac{1}{p} \left( \frac{{\cal \hat{V}}_{IJ}}{{\cal \hat{V}}} - \frac{{\cal \hat{V}}_I {\cal \hat{V}}_J}{{\cal \hat{V}}^2}  \right) \;, \label{hesse_identity3}
\end{equation}
we can use (\ref{hesse_identity1}) and (\ref{hesse_identity2}) to deduce that 
\begin{equation}
a_{IJ} \sigma^J = - {\cal V}_I \;. \label{hesse_identity4}
\end{equation}
It follows that contracting the coordinates with the metric we are left with 
unity:
\begin{equation}
a_{IJ} \sigma^I \sigma^J = 1 \;.  \label{coord_and_dual}
\end{equation}
It is important to note that this is not a constraint on the coordinates 
$\sigma^I$ but an identity which 
follows from the particular form (\ref{hesse_prepotential}) 
of the Hesse potential. As is evident from (\ref{hesse_identity3}) the
metric coefficients $a_{IJ}$ are homogeneous of degree $-2$.
Thus the metric (as a tensor) is homogeneous of degree $0$. As a consequence,
re-scalings $\sigma^I \rightarrow \lambda \sigma^I$ of the affine
coordinates act as isometries on $M$, and also on $N=TM$. This additional
symmetry will be helpful in solving the equations of motion.

We now motivate the introduction of dual coordinates by first noting that 
the equation of motion (\ref{sig_model_eom1}) simplifies if we can find dual
coordinates $\sigma_I$ which satisfy 
\begin{equation}
\dot{\sigma}_I = a_{IJ}(\sigma) \dot{\sigma}^J \;. 
\label{dual_vectors}
\end{equation}
For extremal black holes, where $c=0$, this allows one immediately to
express the solution in terms of harmonic functions, even if no
spherical symmetry is imposed \cite{Mohaupt:2009iq}. 
If $a_{IJ}$ is Hessian, then dual 
coordinates can always be found explicitly and 
are given by $\sigma_{I} \propto {\cal V}_I$. 
From the identity (\ref{hesse_identity4}) we see that these coordinates can 
be written as
\begin{equation}
\sigma_I = -a_{IJ} \sigma^J \;.
\label{dual_coords}
\end{equation}

The minus sign might be counter-intuitive, but one should remember that
the $\sigma^I$ are functions (local coordinates) and not vector fields. 
The dual coordinates $\sigma_I$ 
are algebraic functions of the affine coordinates $\sigma^I$.

For example, if the prepotential is a general homogeneous polynomial 
${\cal \hat{V}} = C_{I_1 \ldots I_p}\sigma^{I_1}\ldots \sigma^{I_p}$ of degree $p$,
then dual coordinates are given by 
\begin{equation}
\sigma_{I} = -\frac{1}{p}\frac{\partial_I C_{I_1 \ldots I_p}\sigma^{I_1}\ldots \sigma^{I_p}}{C_{I_1 \ldots I_p}\sigma^{I_1}\ldots \sigma^{I_p}} \;.
\end{equation}
A special case of particular interest is if the prepotential is of the 
form ${\cal \hat{V}} = \sigma^1 \ldots \sigma^p$ in which case dual 
coordinate are 
\begin{equation}
\sigma_I = -\frac{1}{p}\frac{1}{\sigma^I} \;.
\end{equation}

While it is always possible to find explicit expressions for the dual
coordinates in terms of the affine coordinates $\sigma^I$, 
inverting this relation 
amounts to solving $n$ coupled algebraic equations, which in general cannot
be done in closed form. Solving these equations is in fact equivalent 
to solving the (five-dimensional) black hole attractor 
equations \cite{Mohaupt:2009iq}.

%%%%%%%%%%%%%%%%%%%%%%%%%%%%%%%%%%%%%%%%%%%%%%%%%%%%%%%%%

\subsection{Four-dimensional instanton solutions}

We now proceed to solving the equations of motion (\ref{sig_model_eom1}), (\ref{sig_model_eom2}) and (\ref{sig_model_eom3}). Since they were derived from the 
action of a Euclidean non-linear sigma model, the solutions will be referred
to as 
instantons. We will consider Hessian manifolds of the form (\ref{hesse_prepotential}) and we will formulate the solutions in terms of the dual coordinates, 
making use of the identities derived in the previous section.

The equations of motion (\ref{sig_model_eom2}) for the axions $b^I$ 
are solved by
\begin{equation}
a_{IJ}(\sigma) \dot{b}^I = \tilde{q}_I = \mbox{const.}
\;, \label{axion_charge}
\end{equation} 
where $\tilde{q}_I$ are the `axion charges' 
(or `instanton
charges'), which are the conserved charges corresponding to the
isometries $b^I \rightarrow b^I + C^I$.

Now we turn our attention  to (\ref{sig_model_eom1}). Using the dual
coordinate $\sigma_I$, this becomes
\begin{equation}
\label{eom3v2}
\ddot{\sigma}_I - \frac{1}{2} \partial_I a_{JK}(\sigma) \left( \dot{\sigma}^J \dot{\sigma}^K - \dot{b}^J \dot{b}^K \right) = 0  \;,
\end{equation}
and using that $\partial_I a_{JK} = - a_{JL} a_{KM} \partial_I a^{LM}$ this
can be written as
\begin{equation}
\label{eom3v3}
\ddot{\sigma}_I - \frac{1}{2} \partial_I a^{JK}(\sigma) \left( \dot{\sigma}_J \dot{\sigma}_K - \tilde{q}_J \tilde{q}_K \right) = 0  \;,
\end{equation}

In the extremal case, where the geodesic curve on $N$ is null,\footnote{To be 
precise, the geodesic curve corresponding to an extremal solution 
is not only null, but satisfies $\dot{\sigma}^I = \pm \dot{b}^I$. 
See \cite{EucIII,Mohaupt:2009iq} for an interpretation 
in terms of the para-K\"ahler 
geometry of $N$.} the 
second term is absent, and the equations collapse to $\ddot{\sigma}_I=0$,
which is solved by 
\[
\sigma_I(\tau) = A_I + B_I \tau \;.
\]
In the extremal case the standard radial coordinate (centered at the
horizon) is $\rho$, where $\rho^2 = \frac{1}{2\tau}$, so that \footnote{Affine coordinates are only unique up to affine transformations. The normalization has been chosen for later convenience.}
\[
\sigma_I(\rho) = A_I + \frac{2B_I}{\rho^2} \;.
\]
Thus the solution can be expressed in terms of $n$ spherically symmetric 
harmonic functions, which depend on $2n$ parameters. 
For $c\not=0$ the equation (\ref{eom3v3}) is more complicated
and involves the Christoffel symbols of $N$. To simplify the problem,
we contract (\ref{eom3v3}) with $\sigma^I$, to obtain 
a single equation. This leads to an
enormous simplification, provided that we make full use of the
special properties of the scalar metric. Since the metric 
is homogeneous of degree $-2$, we have 
\[
\sigma^I \partial_I a_{JK} = - 2 a_{JK} \;.
\]
Combining this with (\ref{dual_coords}),
the contracted equations reduces to
\begin{equation}
a^{IJ} \sigma_I \ddot{\sigma}_J = 4c^2 \;. \label{sig_2nd_order} 
\end{equation}
Comparing to the Hessian identity ({\ref{coord_and_dual}}) we 
see that this equation 
implies that 
\begin{equation}
\label{SolContractedEq}
4c^2 \sigma_I = \ddot{\sigma}_I + X_I\;,
\end{equation}
where $X_I$ vanishes when contracted with $\sigma^I$, $\sigma^I X_I =0$.
One obvious strategy is to look for solutions where $X_I=0$.
In this case the equations reduce to the linear equations 
\begin{equation}
4c^2 \sigma_I = \ddot{\sigma}_I \;,
\label{sig_2nd_order_solution}
\end{equation}
which are elementary to solve. We can write the general solution as
\begin{equation}
\label{GenSol}
\sigma_I = A_I \cosh 2c\tau + \frac{1}{2c} B_I \sinh 2c\tau \;,
\end{equation}
where we have chosen the appropriate factors so that in the extremal limit
\begin{equation}
\label{GenSolExLim}
\xymatrix{
\sigma_I \ar[r]_-{c \rightarrow 0} & A_I + B_I \tau \;.
%\sigma_I \ar[r]_(.3){c \rightarrow 0} & A_I + B_I \tau \;.
}
\end{equation}
The solution contains $2n$ arbitrary constants, which is as many as
we expect for the general solution of the original
equation (\ref{eom3v2}). However, we have assumed
without justification that $X_I = 0$, and therefore 
we still have to investigate whether   
(\ref{GenSol}) is a solution, or even 
the general solution, of (\ref{eom3v2}).
Therefore we substitute (\ref{GenSol}) back into
(\ref{eom3v2}). Using $\ddot{\sigma}_I=4c^2 \sigma_I$,
together with 
\[
\sigma_I = -a_{IJ} \sigma^J = \frac{1}{2} \sigma^K \partial_K a_{IJ} \sigma_J
= \frac{1}{2} \partial_I a_{JK} \sigma^J \sigma^K 
= - \frac{1}{2} \partial_I a^{JK} \sigma_J \sigma_K \;,
\]
which combines various of the special identities satisfied by 
$a_{IJ}$, we obtain
\begin{equation}
\label{Constraint1}
\partial_I a^{JK} ( 4c^2 A_J A_K - B_J B_K + \tilde{q}_J \tilde{q}_K ) = 0 \;.
\end{equation}

This equation is to be viewed as an algebraic constraint on 
the integration constants $A_I$ and $B_I$. Since we assume
that the solution for $\sigma_I$ is given by (\ref{GenSol}),
the `Christoffel symbols' $\partial_I a^{JK}$ are functions 
of the integration constants $A_I$, $B_I$ and of the curve
parameter $\tau$. Thus we obtain $n$ algebraic relations
between the $3n$ constants $A_I$, $B_I$ and $\tilde{q}_I$
which have to be satisfied along the geodesic curve, i.e. 
for all values of the curve parameter $\tau$. 
These conditions are hard to investigate without specifying
the scalar metric $a_{IJ}$ explicitly. However, we will prove 
the following three statements in the following sections:
\begin{enumerate}
\item
If the metric $a_{IJ}$ and the Christoffel symbols are 
diagonal (or can be brought to 
diagonal form by a linear transformation of the affine 
coordinates $\sigma^I$), then (\ref{GenSol}), with 
$2n$ independent constants $A_I, B_I$ is the general
solution. In this case the metric of the scalar manifold $N$
is the product of $n$ two-dimensional metrics, and the 
scalars $\sigma_I$ completely decouple from one another.
In the resulting solution all scalars
$\sigma_I$ are independent, in the sense that all mutual
ratios are non-constant, and the corresponding 
five-dimensional
scalars are non-constant. The Reissner-Nordstrom solution is
recovered by taking the five-dimensional scalars to be constant,
which is equivalent to taking all four-dimensional scalars to be
proportional to one another. 
\item
For arbitrary $a_{IJ}$ there is always a solution of the form 
(\ref{GenSol}) depending on 
$n+1$ independent parameters, which can be taken to be the 
charges $\tilde{q}_I$ and the non-extremality parameter $c$.
For these solutions the four-dimensional scalar fields 
$\sigma_I$ are proportional to one another, and the five-dimensional
scalars are constant. The metric is the five-dimensional
Reissner-Nordstrom metric. These solutions   
are therefore non-extremal 
deformations of `double extreme' five-dimensional black holes,
which are extremal black holes with constant (five-dimensional)
scalars.  This result
is not unexpected, but reassuring, because it shows us how to recover the
non-extremal Reissner-Nordstrom solution, with the slight
generalization that we have $n$ independent gauge fields and
thus $n$ independent charges. We call this solution, which can 
be found for all models, the universal solution.
\item
If the metric and the Christoffel symbols are block diagonal,
with $1<k<n$ blocks, or if they can be brought  to this form
by a linear transformation of the affine coordinates $\sigma^I$,
then we obtain solutions of the form (\ref{GenSol}) with $n+k$
independent integration constants. In this case only  
the ratios between four-dimensional
scalars which belong to the same block have to be constant,
and the five-dimensional solutions have $k-1$ parameters
which correspond to changing the values of the scalars at
infinity.
Such block diagonal models provide
intermediate cases between the diagonal models $k=n$ and the
generic models where $k=1$. 
\end{enumerate}

We stress that we of course expect that the general solution 
always has $2n$ independent integration constants, irrespective
of the form of the scalar metric. However,
solutions of the form (\ref{GenSol}), which are obtained by
assuming $X_I=0$, seem only to account for a subset of solutions
if the Christoffel symbols are not diagonizable. The study of
more general solutions is left to future work.

\section{Dimensional Lifting and black hole solutions}

We now proceed to discuss the three cases in turn.

\subsection{The general solution for diagonal models}

Instead of solving (\ref{Constraint1}), we can impose the
stronger condition
\begin{equation}
\label{StrongConstraint}
4c^2 A_J A_K - B_J B_K + \tilde{q}_J \tilde{q}_K = 0 \;.
\end{equation}
If we do not make assumptions on the structure of $a_{IJ}$, 
this has to be true for all values of $J,K$, in order to
solve (\ref{Constraint1}). This imposes severe constraints
on the constants $A_I, B_J$, which, in general, only allows
solutions where all four-dimensional scalars are proportional
to one another. This solution, which we call the universal
solution, will be discussed in the next section. 

In this section we will restrict the scalar metric in such a
way that we obtain the general solution. Specifically,  
we assume that $\partial_I a^{JK} =0$ for $J\not=K$. 
Such models will be referred to as diagonal models in the following. 
For diagonal models (\ref{Constraint1}) is already solved if we
impose (\ref{StrongConstraint}) for $J=K$:
\begin{equation}
\label{WeakConstraint}
4c^2 A_J^2 - B_J^2 + \tilde{q}_J^2 = 0 \;. 
\end{equation}
This equation can be solved explicitly for the $A_I$, or 
for the $B_I$, or for any linear combinations thereof, in 
terms of the charges $\tilde{q}_I$ and of the remaining $n$
independent combinations of the $A_I$ and $B_I$. In the following
it is convenient to 
consider $A_I$ and $B_I$ as independent parameters and to
compute the resulting charges $\tilde{q}_I$ from 
(\ref{WeakConstraint}):
\begin{equation}
\label{qtildeAB}
\tilde{q}_J^2 = B_J^2 - 4 c^2 A_J^2 \;.
\end{equation}

In order to bring the solution to a form suitable for dimensional
lifting and interpretation as a black hole solution, we remember 
that the four-dimensional Euclidean line element takes the form 
of the time-reduced five-dimensional Reissner-Nordstrom metric
(\ref{4D_RN_metric}), irrespective of the details of the matter sector.
Therefore it is natural to replace the radial coordinate $\tau$, which
is an affine parameter for curve in $N$ corresponding to the solution,
by the standard radial coordinate (\ref{eq:tau_r_relation}):
\[
\rho^2 = \frac{c e^{2c\tau}}{\sinh(2c\tau)}  \;.
\]
Observe that in the extremal limit $c\rightarrow 0$ we recover the
relation
\begin{equation}
\label{RhoExLim}
\rho^2 = \frac{1}{2\tau} \;.
\end{equation}
It is useful to note that
\[
\sigma_I = \frac{1}{2e^{-2c\tau}} \left( A_I(1 + e^{-4c\tau}) + \frac{1}{2c}B_I(1 - e^{-4c\tau}) \right) \;.
\]
As discussed earlier the non-extremal Reissner-Nordstrom solution is
obtained from the extremal one through dressing the line element
by the additional harmonic function 
\[
W(\rho) = 1 - \frac{2c}{\rho^2} = e^{-4c\tau} \;.
\]
We now observe that
\[
\sigma_I(\rho) = \frac{H_I(\rho)}{W(\rho)^{1/2}}  \;,
\]
where
\[
H_I(\rho) = A_I + \frac{B_I - 2cA_I}{2\rho^2}
\]
are harmonic functions. Since the extremal solution is given 
by \cite{Mohaupt:2009iq}
\[
\sigma_I^{(\rm extr)} = H_I (\rho) = A_I + \frac{q_I}{\rho^2} \;,
\]
with constants $A_I$ and $q_I$, 
we see that the non-extremal solution is obtained from the extremal
one by dressing the solution  by the additional factor $W^{1/2}(\rho)$.
In addition, the relation between the standard radial coordinate
$\rho$ and the affine parameter $\tau$ depends on $c$ according to
(\ref{eq:tau_r_relation}).
The constants $A_I$ encode the values of the dual scalars
infinity, and are independent of $c$:
\[
A_I = \sigma_I(\rho \rightarrow \infty) \;.
\]
The constants $B_I$ and $q_I$ are related to one another and 
to the charges $\tilde{q}_I$. In the extremal limit they
only differ by constant factors, and their relation is independent
of the $A_I$:
\[
c=0 \Rightarrow q_I = \frac{1}{2}B_I = \pm \frac{1}{2} \tilde{q}_I \;.
\]
For non-extremal solutions the relations between these three sets
of quantities depend on $c$ and on the $A_I$ according to 
(\ref{qtildeAB}) and 
\[
q_I = \frac{1}{2}(B_I - 2c A_I) \;.
\]
Note that a relation of this form is precisely what we should expect from 
the extremal limit of the general solution (\ref{GenSolExLim}) with the 
change of variables (\ref{RhoExLim}). 
Given these identifications, and using the radial coordinate $\rho$,
the relation between the non-extremal and extremal solution is
given by 
\[
\xymatrix{
\sigma_I = \frac{H_I}{W^{1/2}} \ar[r]_-{c\rightarrow 0} & H_I = 
\sigma_I^{(\rm extr)} \;,
}
\]
where $H_I(\rho)$ and $W(\rho)$ are spherically symmetric harmonic
functions in four dimensions.

Our solution depends on $2n+1$ independent parameters: the
values $A_I$ of the scalars at infinity, the non-extremality 
parameter $c$ and the instanton charges $\tilde{q}_I$. 
Instead of the charges $\tilde{q}_I$ we could use alternatively
the integration constants $B_I$ or $q_I$. So far the charges 
$\tilde{q}_I$ are the most natural choice, as they have a direct
physical interpretation as the conserved charges associated
with the axionic shift symmetries. In the extremal limit, the $B_I$ 
and $q_I$ become proportional to the charges $\tilde{q}_I$, 
but in the non-extremal 
case their relation to the $\tilde{q}_I$ is a function of $c$ and
depends on the values $A_I$ of the scalars at infinity. Below
we will see that $q_I$ have a physical interpretation from 
the five-dimensional point of view.

We can lift our solution to five dimensions and control the
extremal limit. 
Since $\sigma^I = - a^{IJ}(\sigma) \sigma_J$, it suggests itself
to define functions $H^I$ by
\[
\sigma^I = W^{1/2} H^I \;.
\]
Note that $H^I H_I = \sigma^I \sigma_I =1$, and due to the
scaling properties of the metric we have
\[
H^I = - a^{IJ}(H) H_J \;.
\]
While the $H_I$ are harmonic functions, the $H^I$ 
are not. However, since the extremal solution is given 
by $\sigma_{I}^{(\rm extr)} = H_I$, the $H^I$ are the solutions
for the scalars $\sigma^I$ in the extremal limit, 
$\sigma^I_{(\rm extr)} = H^I$. 
Thus the above rescaling allows us
to write the non-extremal solution as a rescaled version of the
extremal one, both in terms of the scalars $\sigma^I$ and the
dual scalars $\sigma_I$. 

We now use that the four-dimensional Euclidean metric is
(\ref{RN4d})
\[
ds_4^2 = W^{-1/2} d\rho^2 + W^{1/2} \rho^2 d \Omega^2 \;,
\]
and that the four- and five-dimensional line elements
are related by (\ref{5dKK4d})
\[
ds_5^2 = - e^{2\tilde{\sigma}}dt^2 + e^{-\tilde{\sigma}} ds_4^2 \;, 
\]
where the Kaluza-Klein scalar $\tilde{\sigma}$ is given in terms
of the four-dimensional scalars by
\[
e^{p\tilde{\sigma}} = \hat{\cal V}(\sigma) = W^{p/2} \hat{\cal V}(H) \;.
\]
Therefore the five-dimensional line element takes the form
\begin{eqnarray}
ds_5^2 &=& - W \hat{\cal V}(H)^{2/p} dt^2 + 
\frac{1}{W^{1/2} \hat{\cal V}^{1/p}(H) } \left( \frac{d\rho^2}{W^{1/2}} 
+ W^{1/2} \rho^2 d \Omega^2 \right) \nonumber \\
&=& - W \hat{\cal V}(H)^{2/p} dt^2 + \frac{1}{\hat{\cal V}(H)^{1/p}}
\left( \frac{d \rho^2}{W} + \rho^2 d \Omega^2\right) \;.\nonumber 
\end{eqnarray}
We observe that the five-dimensional Reissner-Nordstrom metric is recovered if
\[
\hat{\cal V}(H)^{1/p} = \frac{1}{\cal H} \;,
\]
where ${\cal H} = 1 + \frac{q}{\rho^2}$. We will see below how
this arises as a particular limit of general solution for
diagonal models.

Remember that we have obtained the general solution by making
the assumption that the model is `diagonal', in the sense that
the Christoffel symbols $\partial_I a^{JK}$ are diagonal in $J,K$  
for all $I$. One class of prepotentials which leads to such models
is 
\[
\hat{\cal V}(\sigma) = \sigma^1 \sigma^2 \cdots \sigma^p \;.
\]
For $p=3$ we recover the five-dimensional STU model, while for 
$p>3$ the resulting models are not supersymmetric, but have 
properties similar to the STU models as far as black hole solutions are
concerned \cite{EucIII,Mohaupt:2009iq}.  
The scalar manifolds $N$ of the four-dimensional models obtained by 
reduction over time are of the form
\[
N = \left( \frac{SU(1,1)}{SO(1,1)} \right)^p \;.
\]
For $p=3$ we obtain the Euclidean version of the four-dimensional
STU model \cite{EucIII,Mohaupt:2009iq}. 

With this choice of prepotential the dual coordinates are
\[
\sigma_I = \frac{1}{\sigma^I} \;.
\]
This can be solved for the original scalars $\sigma^I$, so that
we obtain the solution in closed form:
\[
\sigma^I = \frac{W^{1/2}}{H_I} \;.
\]
Therefore 
\[
\hat{\cal V}(\sigma) = W^{p/2} (H_1 \cdots H_p)^{-1} \;,
%\prod_{I=1}^p H_I^{-1} \;,
\]
and the resulting five-dimensional line element is
\[
ds^2_{(5)} = 
- \frac{W}{(H_1 \cdots H_p)^{2/p}}   dt^2 
+ \left(H_1 \cdots H_p\right)^{1/p}  
\left( \frac{d \rho^2}{W} + \rho^2 d \Omega^2\right) \;.
\]
The non-extremal five-dimensional Reissner-Nordstrom metric
is obtained in the special case where all the harmonic 
functions $H_I$ are proportional to one another:\footnote{The overall
normalization of ${\cal H}$ is fixed by imposing that the five-dimensional
line element approaches the standard line element of five-dimensional
Minkowski space.}
\[
H_1 \propto H_2 \propto \cdots H_p \propto {\cal H}= 1 + \frac{q}{\rho^2} \;,
\]
so that
\[
H_1 \cdots H_p  = {\cal H}^p \;,
\]
and 
\[
ds_{(5)}^2 = - \frac{W}{ {\cal H}^2}  dt^2 + {\cal H}
\left[ W^{-1} d\rho^2 + \rho^2 d\Omega^2_{(3)}\right] \;.
\] 
We can also find explicit expressions for the five-dimensional 
scalars. 
Remember that (\ref{5dScalars4d})
\[
h^I = e^{-\tilde{\sigma}} \sigma^I \;.
\]
Therefore 
\begin{equation}
\label{h-solution}
h^I = \hat{\cal V}(\sigma)^{-1/p} \sigma^I =
\hat{\cal V}(H)^{-1/p} H^I = 
\frac{(H_1 \cdots H_p)^{1/p}}{H_I} \;.
%\frac{ \prod_{K=1}^p H_K^{1/p} } {H_I}
\end{equation}
We observe that $W$ has cancelled out so that we obtain 
the same solution for $h^I$ as in the extremal case \cite{Mohaupt:2009iq}. 
Taking all harmonic functions to be proportional to one
another amounts to taking the five-dimensional scalars to be 
constant. In this case the metric takes the Reissner-Nordstrom
form, as it must. The only difference between this solution 
and Reissner-Nordstrom solution of five-dimensional Einstein 
Maxwell theory is that our solutions are charged under an 
arbitrary number $n$ of abelian gauge fields. 

Our observation that the solution for the five-dimensional scalars
remains the same as in the extremal case raises the question what
happens to the attractor mechanism. As we have discussed previously,
the function $W$ changes sign at $\rho^2 =2c$, and the four-dimensional
metric (\ref{RN4d}) becomes imaginary. For the Reissner-Nordstrom
solution, which we recover by taking all harmonic functions to be
proportional, this corresponds to crossing the outer horizon into
the region where the coordinate $\rho$ becomes time-like. While 
our construction of the solutions via dimensional reduction over
time is a priori only valid for $\rho^2 > 2c$, we know that the
Reissner-Nordstrom solution is obtained by continuing the solution 
to $0<\rho^2 < 2c$ and lifting. Since our general solution can 
be viewed as deforming the Reissner-Nordstrom solution by 
turning non-constant scalar fields, we should expect that 
the general solution remains valid too. To show this we need
to make the assumption that $\hat{\cal V}(H) \not=0$ for $\rho^2>0$
to exclude additional singularities of the line element. 
In the extremal case it is well known that such singularities
are related to scalars field running off to infinity on $\hat{M}$,
or approaching a singular locus of $\hat{M}$ \cite{MayerMohaupt}. This behaviour
can be avoided by choosing suitable initial conditions for the
scalar fields at infinity. In particular, as long we stay 
`close enough' to the Reissner-Nordstrom solution, no additional
singularity can arise. Then the outer and inner horizon are 
still encoded in $W$ and located at $\rho^2 = 2c$ and $\rho=0$,
respectively. Note that while (\ref{RN4d}) becomes imaginary
at $\rho^2=2c$, the resulting five-dimensional remains real because
the Kaluza Klein exponential
\[
e^{\tilde{\sigma}} = W^{1/2} \hat{\cal V}(H)^{1/p}
\]
becomes imaginary, too.\footnote{Here we regard $e^{\tilde{\sigma}}$ as a 
function that becomes imaginary when 
continued to $\rho^2 < 2c$. A more systematic approach would be to replace
$\tilde{\sigma}$ by a new variable. Since $\tilde{\sigma}$ is defined as
the Kaluza Klein scalar for time-like reduction, it is clear that a 
different variable should be introduced  when the reduced dimension 
becomes spacelike. However, we leave a more detailed investigation of
the region between horizons to future work.} 
The overall effect on the five-dimensional
line element is that $\rho$ becomes time-like while $t$ becomes
space-like. We also observe that the solution (\ref{h-solution})
for the five-dimensional scalars is real (and analytical) for
$\rho>0$. Therefore it makes sense to consider the limit $\rho\rightarrow 0$,
which now corresponds to the inner horizon. We find that the
scalars formally exhibit fixed point behaviour, in the sense that the
solution only depends on the charges $q_I$, and becomes independent
of the remaining constants $A_I$:
\[
\xymatrix{
h^I \ar[r]_-{\rho \rightarrow 0} & \frac{(q_1 \cdots q_p)^{1/p}}{q_I} \;.
%\frac{\prod_{K=1}^p q_K^{1/p}}{q_I}
}
\]
However, we need to remember that 
the parameters $q_I$ are different from the 
electric charges $\tilde{q}_I$. 
In order to get a better understanding, let us consider the interpretation 
of the various parameters from the five-dimensional point of view. 
The $\tilde{q}_I$ are, up to normalization, the electric charges 
of the black hole, i.e. the Noether charges associated with the
conserved current $j_{I|\hat{\nu}} = 
\partial^{\hat{\mu}} (a_{IJ}(h) F^J_{\hat{\mu} \hat{\nu}})$. 
The parameters $A_I$ are the values of the four-dimensional dual scalars
$\sigma_I$ at infinity. In five dimensions, these degrees of freedom 
reorganise themselves into $n-1$ scalars and one degree of freedom 
residing in the metric. In our parametrization, the $n$ scalars $h^I$
are subject to the constraint $\hat{\cal V}(h)=1$, and the 
Kaluza-Klein scalar is given by $e^{p\tilde{\sigma}} = \hat{\cal V}(\sigma)$.
For a five-dimensional black hole solution we should normalize 
the metric such that it approaches the five-dimensional Minkowski metric
at infinity:\footnote{Changing this normalization by a constant factor amounts
to rescaling the five-dimensional Newton constant.}
\[
\xymatrix{
e^{\tilde{\sigma}} \ar[r]_-{\rho \rightarrow \infty} & 1 \;.
}
%\rightarrow_{\rho \rightarrow \infty} 1
\]
This imposes one constraint between the constants $A_I$ which reflects
that there are only $n-1$ five-dimensional scalars for which we can
choose asymptotic values. Thus the five-dimensional solution only
depends on $2n$ independent parameters. The additional parameter 
which we gain by dimensional reduction can be interpreted as the
size of the dimension we reduce over, or, equivalently, as the 
ratio between the five-dimensional and four-dimensional Newton 
constant, since
\[
\frac{1}{G_N^{(4)}} = \frac{1}{G_N^{(5)}} \int_0^{2\pi R} dt
\sqrt{|g_{tt}|} = \frac{1}{G_N^{(5)}} 2 \pi R e^{\tilde{\sigma}(\infty)} 
\;.
\]
While we can use natural units and set $G_N^{(5)} = \frac{1}{16\pi}$,
the ratio of $G_N^{(5)}$ and $G_N^{(4)}$ becomes a physical parameter
once we reduce. However this parameter is irrelevant as far as
five-dimensional black holes are concerned. 

The parameters $q_I$ arise as integration constants for the
solution when using the coordinate $\rho$. Their relation
to the electric charges depends on $c$ and the asymptotic scalar
fields through
\[
2 q_I = \sqrt{\tilde{q}_I + 4c A_I^2} - 2 c A_I  \;.
\]
From the five-dimensional point of view the $q_I$ have a direct
physical interpretation because they determine the asymptotics
of the five-dimensional scalars at the inner horizon.

Since $A_I$ (subject to one constraint), $c$, and $q_I$ are
a set of 2n independent parameters, one might say that we have
fixed point behaviour at the inner horizon in the sense that
the scalars become independent of $A_I$ and $c$ and are completely
determined by the $q_I$. While this is formally correct, it is more
natural to consider $A_I$ (subject to one constraint), $c$, and
the charges $\tilde{q}_I$ as the independent parameters. Then the
asymptotic values of the scalars at the inner horizon do depend 
on their values at infinity, and on $c$, in addition to the charges,
but only through $n$ independent combinations $q_I=q_I(\tilde{q}_I, A_I ,c)$.
One might call this a `dressed attractor', or `dressed fixed 
point'.\footnote{We refrain from calling this a `fake attractor.'}
In the extremal limit $q_I$ and $\tilde{q}_I$ become proportional and
the usual attractor behaviour is recovered.

In the extremal case the asymptotic metric at the event horizon is
of Bertotti-Robinson type, hence a product of maximally symmetric
spaces and therefore an alternative ground state. This is not the
case for non-extremal black holes. We also note that the metric 
at the inner horizon has a two-fold dependence on parameters other than
the charges $\tilde{q}_I$: First it depends on $c$ and $A_I$ 
through the $q_I$, second it acquires an additional universal dependence
on $c$ through the additional  harmonic function $W$.

Having identified $A_I= \sigma_I(\infty)$ and $\tilde{q}_I$ or 
$q_I$ as the physical parameters, let us summarize the relation 
between the charges $\tilde{q}_I$, which are the electrical charges
as defined by current conservation in (super-)gravity
and the charges $q_I$ which govern the asymptotics on the inner
horizon,
\[
\tilde{q}_I = 2 \sqrt{ q_I^2 + 2 c q_I \sigma_I(\infty)}  \;,
\]
and the inverse relation:
\[
q_I = \frac{1}{2} \sqrt{ \tilde{q}_I^2 + 4c^2 \sigma_I(\infty)^2 }
- c \sigma_I(\infty) \;.
\]

%\textbf{
We now turn our attention to the outer horizon, which is located at $\rho = \sqrt{2c}$. On the outer horizon the harmonic functions $H_I$ take the values
\[
H_I = \frac{\bar{q}_I}{2c} \;,
\]
where the $\bar{q}_I$ bare a striking relationship to the dressed charges $q_I$ of the inner horizon
\[
\bar{q}_I = \frac{1}{2} \sqrt{ \tilde{q}_I^2 + 4c^2 \sigma_I(\infty)^2 }
+ c \sigma_I(\infty) \;.
\]
Inspection of the scalar fields on the outer horizon reveals the 
limit\footnote{For completeness we remark that a similar relation holds at radius $\rho=\sqrt{c}$, with $\bar{q}_I$ replaced by the integration constants $B_I$.}
\[
\xymatrix{
h^I \ar[r]_-{\rho \rightarrow \sqrt{2c}} & \frac{(\bar{q}_1 \cdots \bar{q}_p)^{1/p}}{\bar{q}_I} \;.
}
\]
We can interpret the $\bar{q}_I$ as dressed charges which determine the values of the scalars on the outer horizon. 
In this sense they exhibit similar `dressed attractor' behaviour on the outer 
horizon as on the inner horizon. In particular, we observe formally the same 
fixed point behaviour in the extremal limit. Indeed, this must be the case as 
in the extremal limit the outer and inner horizons coincide. The dressed charges on the inner and outer horizon are related to the electric charges through
\[
\tilde{q}_I^2 = 4 q_I \bar{q}_I \;.
\]
%}
While the `dressed attractor' behaviour is not attractor behaviour 
in the proper sense, it demonstrates that the functional dependence
of the solution on the integration constants is not generic, but
takes a restricted form. This is what one should expect if the field
equations can be reduced to first order form.

One important feature of non-extremal charged solutions is  
that the coordinate $\rho$ becomes time-like at the outer
horizon. Therefore the flow becomes a flow in time rather than 
in space between the outer and inner horizon. This should have interesting
implications for time-dependent solutions and in particular cosmology, 
since the between horizon region of non-extremal black holes is a 
natural starting point for the construction of (S-brane type) cosmological
solutions \cite{Behrndt:1994ev,Gutperle:2002ai}. A related question 
is whether something can be learned about the time evolution 
of non-extremal black holes, which are expected to loose mass
through Hawking radiation and to approach the extremal limit.

In the context of string theory,
supergravity provides the macroscopic ($=$ long wavelength $=$ low energy)
description of black holes. For some types of black holes string theory
provides a microscopic description of black holes in terms of strings,
D-branes, and other string solitons. While extremal black holes correspond 
to ground states of brane configurations, non-extremal black holes
correspond to excited states. Since our class of solutions contains
the five-dimensional STU model, which occurs as a subsector in various
string compactifications, it is natural to use these models to 
investigate the microscopic interpretation of our solutions.

\subsection{The STU model and IIB string theory on $T^5$}

The five-dimensional STU model is based on the Hesse potential
\[
{\cal V} = - \log (\sigma^1 \sigma^2 \sigma^3) = -\log \sigma^1 -\log \sigma^2 -\log \sigma^3 \;.
\]
It describes two vector multiplets coupled to supergravity and 
arises (together with hypermultiplets which can be truncated out
consistently) as the classical limit of the
compactification of the heterotic string on 
$K3 \times S^1$ with instanton numbers $(12-n,12+n)$, 
where $n=0,1,2$ \cite{d=5Het}.
Furthermore, it arises as a universal subsector in 
compactifications with $N=4$ and $N=8$ supersymmetry, in particular
in type-II compactifications on $T^5$, as reviewed in 
\cite{Maldacena:1996ky}. 
Let us first collect 
the relevant formulae:
The five-dimensional line element is given as
\begin{align*}
ds_{(5)}^2 	&= -e^{2\tilde{\sigma}}dt^2 + e^{-\tilde{\sigma}} ds^2_{(4)}	\\
&= - \frac{W}{ (H_1 H_2 H_3)^{2/3} } dt^2 + (H_1 H_2 H_3)^{1/3} \left[ W^{-1} d\rho^2 + \rho^2 d\Omega^2_{(3)} \right] \;,
\end{align*} 
and the five-dimensional scalars $h^I$ are given by
\[
h^I = e^{-\tilde{\sigma}} \sigma^I = \left( \frac{ H_J H_K }{ H_I^2 } \right)^{1/3} \;,
\]
where $I,J,K$ are pairwise distinct. The limit on the inner
horizon is:
\[
\xymatrix{
h^I \ar[r]_-{\rho \rightarrow 0} &
\left( \frac{q_J q_K }{ q_I^2 } \right)^{1/3} \;.
}
%\begin{CD}
%h^I @>>{\rho\rightarrow 0}> \left( \frac{q_J q_K }{ q_I^2 } \right)^{1/3} \;.
%\end{CD}
\]
The same solution was found in \cite{Maldacena:1996ix,Maldacena:1996ky}, 
using
the results of \cite{MahSchw}, by compactification
of the type-IIB string theory on $T^5$. One particular realization 
is a system which carries integer D1-brane charge $Q_1$, 
integer D5-brane charge $Q_5$ and integer momentum $N$ along
the D1-brane. These charges can be expressed in terms of 
the string coupling $g$, the radii $R_5, \ldots, R_9$, 
the non-extremality parameter $c$ and three `boost parameters'\footnote{
The original notation in \cite{Maldacena:1996ky} is $\alpha, \gamma,\sigma$,
and $Q_N$ is denoted $N$. 
Also note that in comparison to \cite{Maldacena:1996ky} $r_0^2 = 2c$.}
$\alpha_1, \alpha_5, \alpha_N$ as follows:
\[
Q_1 =  \frac{V}{g} c \sinh(2\alpha_1) \;,\;\;\;
Q_5 =  \frac{1}{g} c \sinh(2\alpha_5) \;,\;\;\;
Q_N= \frac{R^2V}{g} c \sinh(2 \alpha_N) \;,
\]
where $V=R_5 R_6 R_7 R_8$ and $R=R_9$. Since the underlying brane
configuration consists of D1 branes oriented along the $x_9$ direction
within the D5 world volume, the moduli are the radius $R=R_9$, 
the volume $V$ of the torus spanned by the other four compact directions,
and the string coupling.
%\footnote{Note that in \cite{Maldacena:1996ky}
%the harmonic functions are normalized to 1 at infinity. This has the
%effect that the moduli are not encoded in the constant part of the
%harmonic function, but appear as `dressing factors' in front of the
%charges.}

In \cite{Maldacena:1996ky} the extremal limit is performed by
sending $c\rightarrow 0$, and the boost parameters $\alpha_I \rightarrow
\infty$, while keeping the brane charges $Q_I$ and the moduli
$g,R,V$ constant. 

To relate this to our solutions, we note that harmonic functions
in \cite{Maldacena:1996ky} take the form
\[
H_I = 1 + \frac{2c \sinh \alpha_I}{\rho^2} \;.
\]
Matching this with our parametrization\footnote{We now let the indices
$I$ take values $I=1,5,N$ instead of $1,2,3$.}
\[
H_I = \sigma_I(\infty) + \frac{q_I}{\rho^2}\;,
\]
we observe  that in \cite{Maldacena:1996ky} the constant 
terms are normalized to 1, which has the effect that the moduli
dependence is scaled into the $\frac{1}{\rho^2}$ term. 
To understand the relation between the brane charges $Q_I$ and our inner horizon
charges $q_I$ it is sufficient to set
$\sigma_I(\infty)=1$. Then
\[
q_I = c \sinh^2 (\alpha_I) \;, 
\]
and using this we find:
\begin{equation}
Q_1 = 2 \frac{V}{g} \sqrt{ q_I (q_I + c) } \;,\;\;\;
Q_5 = 2 \frac{1}{g} \sqrt{ q_I (q_I + c) } \;,\;\;\;
Q_N = 2 \frac{R^2 V}{g^2} \sqrt{ q_I (q_I + c) } \;.
\end{equation}
Thus for fixed moduli $V,R,g$ the charges $Q_I$ and $q_I$ 
are proportional, up to higher order terms in $c$. 
From the microscopic
point of view it is natural to perform the extremal limit 
such that 
the integer valued charges $Q_I$ are kept fixed. Then 
$q_I$ and $\tilde{q}_I$ are not constant, but the extra 
terms are subleading in $c$.

For completeness we mention that in the non-extremal case
the integer valued charges do not count the total numbers of
D1 branes, D5 branes and quanta of momentum, but the differences
in the numbers of branes and anti-branes, and of left- and right
moving momenta. 
Non-extremal black holes can be interpreted as systems
of branes and anti-branes, and, surprisingly, the resulting
formulae for mass and entropy look like those of a non-interacting
system \footnote{Callan''1996dv,Horowitz:1996fn,Maldacena:1996ky}. 
It should be interesting to investigate whether the
`dressed attractor mechanism' described above can shed some light
onto such systems and, possibly, onto their dynamical evolution 
towards the extremal limit.

\subsection{The universal solution}

Let us now return to the general class of models, where we 
do not make any additional assumptions about the scalar
metric. We can still find a solution by imposing 
(\ref{StrongConstraint})
\[
4c^2 A_J A_K - B_J B_K + \tilde{q}_J \tilde{q}_K = 0 \;,
\]
but in order to solve the original constraint (\ref{Constraint1})
this must now hold for all values for $J$ and $K$. 
Already the equations where
$J=K$ fix $n$ constants. For example we can solve for the
$B_J$ in terms of $A_J$ and the charges $\tilde{q}_I$:
\[
B_J =  \sqrt{\tilde{q}_J^2 + 4 c^2 A_J^2}\;.
\]
The remaining equations, where $J\not=K$, can only be solved
if we take $A_I \propto \tilde{q}_I$, which in turn implies
that $B_I \propto \tilde{q}_I$. The possible solutions can be parametrized in the form
\[
A_J = \mu \tilde{q}_J \;,\;\;\;
B_K =  \tilde{q}_K  \sqrt{1 + 4c^2 \mu^2 }\;, 
\]
where $\mu$ is a parameter which reflects that the overall normalization
of $A_J, B_K$ relative to the charges is not fixed by the constraint.

Writing the solution in the form
\[
\sigma_I(\rho) = \frac{H_I(\rho)}{W^{1/2}(\rho)} \;,\;\;\;
H_I(\rho) = \sigma_I(\infty) + \frac{q_I}{\rho^2}  \;,
\]
we find
\[
A_I = \sigma_I(\infty) = \mu \tilde{q}_I \;,\;\;\;
q_I = \frac{1}{2} \tilde{q}_I \left( \sqrt{1+4c^2 \mu^2} - 2 c \mu \right)
\;.
\]
Therefore the harmonic functions $H_I$ are proportional to one another,
\[
H_1 \propto H_2 \propto \cdots \propto H_p \propto {\cal H} = 
1 + \frac{q}{\rho^2} \;,
\]
and the solution can be expressed in terms of two independent functions
$W(\rho)$ and ${\cal H}(\rho)$. 
This implies that the metric takes the Reissner-Nordstrom form,
and that the five-dimensional scalars are constant. In the 
previous section we derived this for diagonal models, but it remains
valid here because we only need to use the homogeneity properties of the 
sclar metric. Since all harmonic functions are proportional,
we are effectively dealing with homogeneous functions of one variable,
which are determined, up to overall normalization, by their degree.

In particular, the dual scalars $\sigma_I$ are homogeneous functions of degree
$-1$ of the scalars $\sigma^I$. Given that the universal solution 
takes the form $\sigma_I \propto {\cal H} W^{-1/2}$, it follows that 
$\sigma^I \propto {\cal H}^{-1} W^{1/2}$. The prepotential is homogeneous
of degree $p$, and therefore 
\[
\hat{\cal V}(\sigma) \propto W^{p/2} {\cal H}^{-p} \;,
\]
which implies that the line element takes the Reissner-Nordstrom form.
The five-dimensional scalars $h^I$ are homogeneous of degree zero
in the harmonic function, and therefore must be constant if 
the harmonic functions are proportional. 

This also clarifies the role of the parameter $\mu$. When lifting
to five dimensions we impose the normalization condition 
\[
\xymatrix{
e^{\tilde{\sigma}(\rho)} = \frac{W^{1/2}}{\hat{\cal V}(H(\rho))^{1/p}}
\ar[r]_-{\rho \rightarrow \infty} & 1 \;.
}
\]
This is a condition on the asymptotic four-dimensional scalars
$\sigma_I(\infty) = \mu \tilde{q}_I$, which for the universal
solution are proportional to the charges. Therefore the parameter
$\mu$ needs to be used to normalize the five-dimensional metric.
In the four-dimensional set-up, $\mu$ is not fixed and 
encodes the relation between the five-dimensional and four-dimensional
Newton constant.

\subsection{Block-diagonal models}

There are intermediate cases where the Christoffel symbols 
$\partial_I a^{JK}$ simultaneously assume block-diagonal form,
or can brought to this form, by a linear transformation. For 
concreteness, suppose that the indices split into two subsets
\[
1 \leq J_1, K_1 \leq m \;,\;\;\;
m < J_2, K_2 \leq n \;,
\]
such that $\partial_I a^{I_1 J_2} = 0$ for all $I$. 
Then we obtain a solution of (\ref{Constraint1})
by imposing (\ref{StrongConstraint}) for $J=K$, $J_1\not= K_1$ 
and $J_2 \not=K_2$, but we do not need to impose it if $J$ and $K$ 
belong to different blocks. 

The `diagonal' constraints imply
\[
B_J = \sqrt{\tilde{q}_J + 4 c^2 A_J^2 } \;.
\]
But since there are no `off-diagonal' constraints if $I$ and $J$ 
belong to different blocks, we obtain
\[
A_{J_k} = \mu_k \tilde{q}_{J_k} \;,\;\;\;
B_{J_k} = \tilde{q}_{J_k} \sqrt{1 + 4 c^2 \mu_k^2} \;,
\]
where $k=1,2$. As a result only harmonic functions belonging
to the same block must be proportional to one another:
\[
H_1 \propto \cdots H_m \propto {\cal H}_1 \;,\;\;\;
H_{m+1} \propto \cdots H_n \propto {\cal H}_2 \;,
\]
and the solution depends on three independent harmonic
functions $W,{\cal H}_1, {\cal H}_2$. After lifting 
to five dimensions one combination of the parameters 
$\mu_1$ and $\mu_2$ is fixed by normalizing the metric
at infinity. There remains one undetermined parameter which
allows to vary the value of one five-dimensional
scalar field at infinity. 

For models with a larger number of blocks the number of 
undetermined moduli at infinity and hence of non-constant
scalar fields increases. If the Christoffel symbols 
decompose into $k$ blocks, then $k-1$ five-dimensional
scalars can be non-constant. While $k=1$ corresponds 
to the universal solution, where all scalars are constant,
$k=n$ corresponds to diagonal models, where all $n-1$
five-dimensional scalars can be non-constant.

Block-diagonal Christoffel symbols with two blocks occur when the 
Hesse potential takes the form
\[
{\cal V} = - \frac{1}{p} \log \left( {\hat{\cal V}}_1
(\sigma^1, \ldots, \sigma^m)
\hat{\cal V}_2(\sigma^{m+1}, \ldots, \sigma^n) \right) \;,
\]
where ${\cal V}_1$ and ${\cal V}_2$ are homogeneous functions 
of degrees $r$ and $s$, where $r+s=p$.
A higher number of blocks occurs when the Hesse potential 
factorizes into more homogeneous factors, and the extreme 
case of a diagonal model occurs for complete factorization into
factors of degree one, $\hat{\cal V}_I \propto \sigma_I$.

Of course we expect that even for generic models solutions
exist, where all scalars are non-constant,
because such solutions exist in the extremal limit. However
the solutions which we have constructed explicitly in this
article only have a limited
number of non-constant scalar fields. 
Metrics where the prepotential factorizes into independent
homogeneous factors are in particular product metrics and 
therefore rather special. Thus it is important to 
make progress by finding more general solutions for
models without block structure.

\section{Conclusions and Outlook}

In this paper we have demonstrated that 
non-extremal black hole solutions can be obtained
from extremal ones by a universal deformation which 
is blind to the details of the matter sector. 
While the class of models for which 
explicit solutions were obtained happens to be 
based on symmetric spaces,
the relevant features for obtaining solutions were
given by 
the generalized special geometry, through the existence
of a potential together with homogeneity properties.
What played a crucial role, however, was the factorization 
of the target space into two-dimensional spaces with 
simple geodesics, as is clear from the fact that
the number of explicit solutions that we could obtain 
is correlated with 
the number of blocks into which the scalar metric
can be decomposed. Therefore we expect that further
progress will require a more detailed understanding of
geodesics in generalized special real manifolds. 
Since the general analysis of the field equations
allows the presence of an extra term in the
contracted scalar field equation 
(\ref{SolContractedEq}), which vanishes for
diagonal models, this term is likely to come into play for
non-diagonal models. It is encouraging that the geometry
obtained by reducing the black hole space-time with
respect to time, the time-reduced Reissner-Nordstrom
metric, is completely fixed and independent of the
matter sector. The other feature which we observed,
and which works universally in diagonal models, is that
the non-extremal solution is obtained by dressing 
metric and scalar fields by an additional harmonic function. 
Since this is closely related to the homogeneity properties of 
the scalar manifold, which also hold for non-diagonal models,
we expect that progress can be made without assuming that the
target space is a symmetric space. The problem of
solving the field equations amounts to constructing a 
harmonic map from the reduced space time into the target space.
For spherically symmetric solutions this reduces to constructing
geodesic curves. The difference between extremal and non-extremal
solutions is that the former correspond to null geodesics while
the later ones correspond to space-like geodesics. A further
difference, which is obscured by the reduction to the radial 
coordinate, but manifest as long as we only reduce over time, is
that for extremal solutions the time-reduced geometry is flat,
while for non-extremal solutions it is only conformally flat.\footnote{Here
we use that any spherically symmetric metric can be brought to isotropic
form \cite{Kustaanheimo}.} 
This shows how the harmonic map gets deformed when making
solutions non-extremal: the geometry of the reduced space-time
is modified by a conformal factor, which forces the geodesic 
to become non-null, and this manifests itself through the 
dressing of the solution by an additional harmonic function. 
Upon reduction to the radial coordinate the conformal factor
of the reduced space-time becomes encoded in the relation between
the standard radial coordinate $\rho$ and the affine curve parameter 
$\tau$ of 
the geodesic. None of these observations are specific to 
diagonal models, and thus we expect that the general class of
models can be understood by digging deeper into the geometry 
of the harmonic map. 

It should also be instructive to relate our work to approaches
based on first order flow equations and integrability
\cite{CerDal,CCDOP,GNPW:05,GaiLiPadi:08,Perz:2008kh,ChiGut,CRTV,ADOT}. Flow
equations and harmonic functions are intimately related. 
In \cite{Mohaupt:2009iq} the reduction of the harmonic equation to a
first order equation was shown to be the result of the existence
of $n$ conserved charges. While this was done for extremal solutions,
only, the argument should carry over to the non-extremal solutions considered
here, because in terms of the radial variable $\rho$ the solution 
for the five-dimensional scalar fields remains the same. The non-extremal
deformation is fully encoded in the modified relation between the
radial variable $\rho$ and the affine parameter $\tau$. 
This is interesting, because the argument given in \cite{Mohaupt:2009iq}
does not require the target space to be symmetric, but only the
existence of $n$ isometries. The approach via symmetric spaces
is closely related to integrability and the Hamilton-Jacobi
formalism. The latter is used 
in order to identify adapted parametrizations
of the field equations. Our approach uses geometrical considerations
in order to arrive directly at such a parametrization, given by 
the dual scalar fields $\sigma_I$ and the affine curve parameter $\tau$.
For extremal black holes this was briefly investigated in \cite{Mohaupt:2009iq},
and we plan to explore this more systematically in the future.

In this paper we have restricted ourselves to static, five-dimensional
black holes. The extension to various other types of solutions 
should be interesting to investigate. Since supersymmetry does not
play an immediate role, the adaptation of our results to dimensions
other than four is straightforward and amounts to adjusting numerical
factors. However, by working in five dimensions we have restricted 
ourselves to electric charges, while in four dimensions generic charged
black holes carry both electric and magnetic charge. Applying 
dimensional reduction to this case leads to a more complicated 
target space geometry, with an isometry group which is solvable (of 
Heisenberg group type) rather than abelian, as is well known 
from the c-map \cite{FerSab,BGLMM}. 
We believe that this is best approached systematically
by re-visiting and generalizing the c-map, which we leave
to future work. At the current stage we see no problem
in principle, and expect that the features we have observed will 
pertain. 

Other extensions would naturally include the study of rotating solutions,
the addition of a cosmological constant, Taub-NUT charge (i.e.
more complicated Ricci flat and conformally Ricci flat time-reduced
geometries), black strings  and black rings, domain walls 
and cosmological solutions. Of course there is already a large 
literature on all these types of solutions, and dimensional reduction 
is often used as one of the tools. For example, black ring solutions
were constructed using reduction over time in \cite{Yazadjiev}. 
However, we believe that dimensional reduction could play an even
bigger role in particular in handling generic matter sectors and
organizing solutions, if the underlying geometry of harmonic maps 
is fully exploited. Concerning cosmological solutions it
is interesting that we found solutions which extend to the inner
horizon, because the Killing vector becomes space-like between
the horizons. Thus the scalar flow becomes a flow in time between
the horizons. The between-horizon geometry of charged, non-extremal
solution is a natural starting point for the construction of 
cosmological solutions of the S-brane type 
\cite{Behrndt:1994ev,Gutperle:2002ai}. Non-extremal black hole solutions
can also be used to obtain `mirage-type' cosmologies, where FRW cosmology
is induced on branes moving in the black hole background \cite{Cardoso:2008gm}.

It has been
observed that in cases where a reduction of the field equations to
first order flow equations takes place, there is a close relation 
between black holes and other types of solutions including domain 
walls, instantons and cosmologies. The frameworks proposed 
for capturing these relations are characterized by the key words
`fake (super-)potentials', `fake-' or `pseudo-'Killing spinors
and `fake supersymmetry' \cite{Fake1,Fake2}. The
`generalized special geometries' used in this article are similar
in spirit as they also aim to extend techniques
originally developed within a supersymmetric set-up to more general
non-supersymmetric situations.
It should be interesting to explore the relations between
these frameworks.
We note that the reduction over time introduces `variant real forms'
of special geometry, specifically the Euclidean special geometries
described in \cite{EucI,EucII,EucIII}.\footnote{The para-K\"ahler geometry
of the extended scalar space obtained by reduction over time can be viewed 
as a generalization of the projective special para-K\"ahler geometry of
Euclidean four-dimensional $N=2$ vector multiplets constructed in 
\cite{EucIII}. } Similar observations have been made with regard to 
maximal supergravities, their toroidal reductions and the temporal 
T-dualization  of type-II string theories 
\cite{CLLPST,HullJulia,Hull:1998,Bergshoeff:Complex}. 
This indicates 
a unifying pattern underlying (super-)gravity solutions, branes,
and their various mutual relations, which deserves further 
exploration.

\end{document}